\documentclass[12pt, a4paper]{article}
\usepackage{geometry}

\usepackage[square,numbers,sort&compress]{natbib}
\usepackage[utf8]{inputenc}

\usepackage{array}
\usepackage{amsthm}
\usepackage{amsmath}
\usepackage{amssymb}
\usepackage{slashed}
\usepackage{bm}
\usepackage{graphicx}
\usepackage{wrapfig}
\usepackage{physics}
\usepackage{mathtools}
\usepackage{subfiles}
\usepackage{color, colortbl}
\usepackage{hyperref}
\usepackage{marginnote}
\usepackage{caption}
\usepackage{subcaption}
\usepackage{xr}
\usepackage[counter,user]{zref}
\usepackage{float}
\usepackage{authblk}
\usepackage{blindtext}

\newcommand{\inb}[3]{{\left#1 #2 \right#3}}
\newcommand{\reff}[1]{Fig.\ref{fig:#1}}
\newcommand{\refe}[1]{Eq.\ref{eq:#1}}
\newcommand{\Hs}[0]{Hamiltonians}

\title{SPT extension of $Z_2$ quantum Ising model's ferromagnetic phase} %Title of paper

\author[1]{Hrant Topchyan}
\affil[1]{\small Alihkanyan National Laboratory, Yerevan Physics Institute, Yerevan, Armenia}

\begin{document}

\maketitle

\begin{abstract}
This paper focuses on the creation of a model with explicitly defined
symmetry-protected topological (SPT) phases on a triangular lattice
as an extension of $Z_2$ Ising model's ferromagnetic phase.
Unlike in previously known similar works,
this model is based on an initially interacting system
which is known to describe experimentally realizable physical systems.
The Hamiltonian for these edge states contains
four-point spin interactions between next-to-next nearest neighbors.
As an initially interacting
A generic technique for creating SPT models is developed,
allowing for the construction of translation-invariant edge models.
\end{abstract}


\maketitle %\maketitle must follow title, authors, abstract and \pacs

\section{Introduction}
Symmetry-protected topological (SPT) phases
\cite{spt-over,spt-class-1,spt-class-2,spt-class-3}
are one of the hot topics in the current research arena
\cite{levin-gu, spt-1, spt-2, spt-3, spt-4, spt-5, spt-6,
spt-7, spt-8, spt-9, spt-10, spt-11, spt-12, spt-13, spt-14,
spt-15, spt-16, spt-17, spt-18, spt-19, spt-20, our-1, our-2,
spt-cohom-1, spt-cohom-2, spt-cohom-3, spt-cohom-4, spt-cohom-5}
both in terms of conceptual formulation and specific model studies.
It is a relatively new concept for phase transitions
strongly related to the topological properties of the system, that is
significantly different from the classical Landau approach to the topic.

Systems with SPT orders are remotely similar to models
described by the Landau theory of phase transitions.
They are both described by the structure and properties of their symmetry groups,
however, the Landau theory relies on explicitly or spontaneously breaking
of the system's symmetries as the inducer of different phases and
in SPT-ordered systems, the symmetry is not broken in either phase.
SPT phases have a topological origin,
and the corresponding states are manifested on the edge of the system,
similar to topologically ordered states.
However SPT ordered systems are known \cite{spt-cohom-1, spt-short-1}
to be short-range entangled unlike the topologically ordered ones, which are long-range entangled
\cite{top-1, top-2, top-3, top-4, top-5, top-6}.

An important feature of SPT order is its support for
symmetry-protected gapless boundary phases,
meaning it can be a topological insulator or a trivial insulator
in different phases.
This behavior is useful for topological quantum computation.
Other remarkable properties of SPT-ordered systems
are the non-standard excitation algebra that emerges for edge states and
the sensitivity of the system to symmetry-breaking perturbations.
They also carry other interesting characteristics,
that might later prove useful in some applications.

In the meantime, a significant amount of research has been done on this topic.
That revealed the explicit connection between
the symmetry classes and the SPT phases,
which happens to be
\cite{spt-cohom-1, spt-cohom-2, spt-cohom-3, spt-cohom-4, spt-cohom-5}
a representation of the third group cohomology of the symmetry.
This allows classification
\cite{spt-class-1, spt-class-2, spt-class-3}
and a deeper understanding of SPT models and their variety.
However, none of these models provide a precise pathway for formulating
an SPT system model or modifying a known model to SPT phase capable.

An outstanding example of an explicitly written SPT model
is presented by Levin and Gu\cite{levin-gu},
which is based on the Ising paramagnet with $Z_2$ symmetry.
The result is a model with two phases,
one being a conventional insulator (with a gapped spectrum),
and the second being a topological insulator
(with gapless spectrum created by edge states).
There are other such known models\cite{our-1, our-2},
which are basically extensions of the Levin-Gu model\cite{levin-gu} for $Z_3$ and $Z_3 \times Z_3$ symmetries.
This paper is inspired by those works,
and is trying to apply the ideas proposed there
to formulate a similar model that is based on a more complex system
(namely the ferromagnetic phase of $Z_2$ quantum Ising model),
that initially contains interaction,
meanwhile developing a universal technique for doing so.

According to the conventional definition,
SPT models have the following definitive properties.
The system has to have a global symmetry $S$
which is neither explicitly nor spontaneously broken in any of the phases.
There is the so-called "trivial" phase which,
generally speaking, has a gapped spectrum,
and frequently has the simplest form of the Hamiltonian
(it is usually possible to write the ground state as
a direct product of different subsystem states).
The other phases are separated from the trivial one and each other.
The phase separation is implemented as the impossibility of
continuously connecting (with any parametrization)
states of different phases without breaking at least one symmetry
or closing the gap of the spectrum.
So the phases are protected by symmetry.
When considered on a system with a boundary,
the non-trivial phases of SPT models tend to have
gapless edge spectrum.

\begin{figure}
    \centering
    \includegraphics[width=200px]{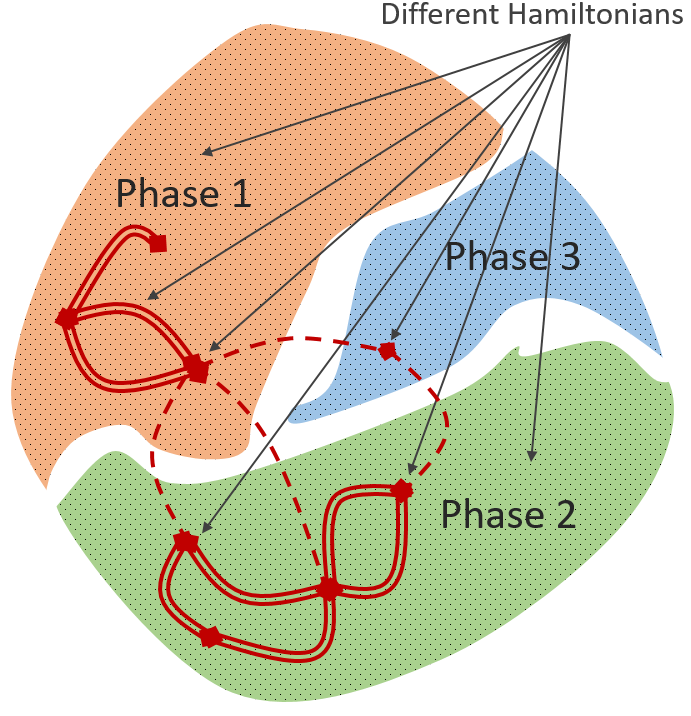}
    \caption{\label{fig:phase_space}Schematic representation of supposed SPT phase space.
        Double lines are continuous connections of {\Hs},
        and dashed lines are non-continuous connections.}
\end{figure}

A phase space is generally defined as a set of {\Hs}
which describes the state of the system in different conditions.
In Ginsburg-Landau theory
those {\Hs} differ by some parameters
(for example, a parameter describing an external magnetic field
or a coupling constant), called critical parameters.
There is also a critical point:
a value of the parameter where the phase transition occurs.

On the contrary, the phase space of SPT models can not be parametrized.
For SPT models, the phase space is a set of gapped bulk {\Hs} with
some common symmetry $S$.
Then a phase is defined as a "region" of those {\Hs},
that can be continuously connected within the set (without breaking the symmetry),
and two {\Hs}, that don't have such a connection
are said to be in different phases.
By the very definition, it is demanded
that within the considered phase space
there is no such thing as a "critical point" in this theory,
as the existence of one would mean continuity of Hamiltonian transformation
inside the set of phase spaces {\Hs}, which are all symmetric.
So the "regions" of different phases don't "touch", as portrayed in \reff{phase_space}.
Of course, it is possible to change the phase continuously,
but this will require leaving the initially considered phase space,
which will either break the symmetry or close the bulk spectrum gap.

The phase properties are defined by the spectrum.
Two {\Hs} having identical spectra (being unitarily equivalent)
is the extreme for defining a phase.
And obviously,
the unitary transformations connecting those {\Hs} should also be symmetric under $S$.
SPT principles suggest, that
there should be no long-range interactions in the system,
so the transformations should be local.
Summing up,
there is a set of continuously symmetric local unitary transformations
that define a single phase.

\section{Construction of SPT models}

The different phases occur as a result of the topology of the system:
there would be no manifestation of different phases
if the system didn't have an edge.
In that case all the {\Hs} would be equivalent,
and all of them would be connected through
an even larger set of continuously symmetric local unitary transformations.
Yet some of them are no longer applicable for edged systems,
as their symmetry might be broken.
This leads to the assumption, that
transitions between different phases are done via
transformations which are related to the unitary transformations
whose symmetry is broken on the edge of the system. 
Our approach is to explicitly 
restore the symmetry of those transformations,
on the expense of their unitarity,
and thus generate the {\Hs} for non-trivial phases,
as it is done in some papers \cite{levin-gu, our-1, our-2}.

The SPT phases are known \cite{spt-class-1, spt-class-2, spt-class-3}
to be classified by cohomologies of their symmetry group, and
the concept of group cohomologies is heavily used
during the construction of a model with SPT phases
\cite{spt-cohom-1, spt-cohom-2, spt-cohom-3, spt-cohom-4, spt-cohom-5}. 

\subsection{Group cohomologies}

\begin{figure}
    \centering
    \includegraphics[width=300px]{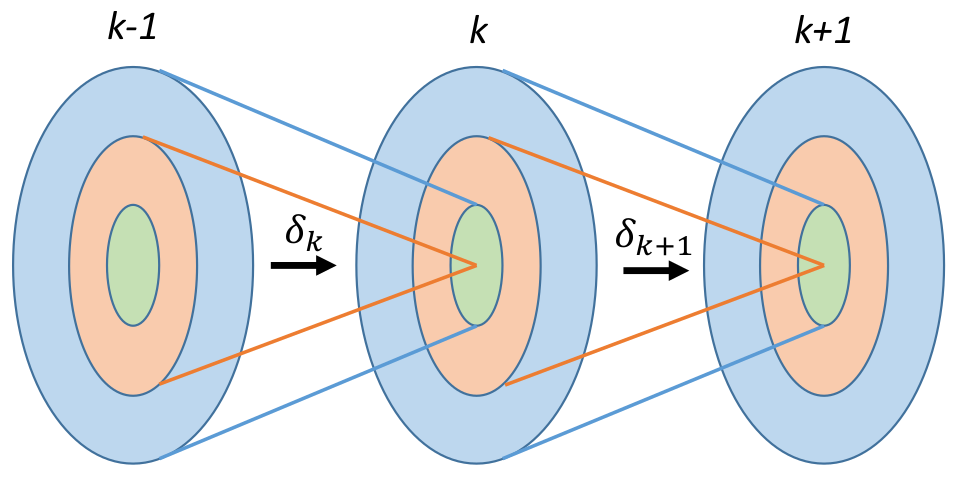}
    \caption{\label{fig:cohoms}Mapping by $\delta$ between $C_k$-s.
        Each set of invested disks represents a cochain space
        with the cochain index specified above.
        The green disks are coboundaries (exact forms),
        the reds are cocycles (closed forms),
        and the blues are cochains (all forms).}
\end{figure}

Group cohomologies are defined by functions
that map multiple parameters of a specified group $G$ to some other group $F$,
and they are symmetric under the generators of $G$ \cite{cohoms-2}.
In other words.
\begin{equation}
    \nu_k : G^{k+1} \rightarrow F, \hspace{1cm}
    \nu_k (g_0, g_1, ... g_k) = \nu_k (0, -g_0 + g_1, ... , -g_0 + g_k )
\end{equation}
Here $g_i$ are any elements of a group $G$ in additive representation.
We will call $\nu_k$ a $k$-cochain or a $k$-form.
Let's denote the space of all cochains by $C_k$.

We can define a so-called coboundary operator on cochains as
\begin{equation}
    \delta \nu_{k-1} (g_0, g_1, ... , g_k) =
    \prod_{i=0}^k {\nu_{k-1} (g_0, g_1, ... , \check{g}_i, ... , g_k)}^{(-1)^i}
\end{equation}
where a "check" on the argument means that the argument is dropped.
For example $\delta \nu_0 (g_0, g_1) =
\frac{\nu_0 (g_1)}{\nu_0 (g_0)} = 1$
because of the symmetric properties of cochains, or
$\delta \nu_1 (g_0, g_1, g_2) =
\frac{\nu_{1} (g_1, g_2) \nu_{1} (g_0, g_1)}{\nu_{1} (g_0, g_2)}$.
As it is portrayed, the coboundary operation maps $C_k$ to $C_{n+1}$
as the number of arguments has increased, but the symmetry is not broken.
The functions that can be generated with $\delta$ acting on $\nu_k$
are called coboundaries or exact forms (we will see why in a minute).

It can be easily shown, that like the geometric boundary operator, $\delta^2$ is trivial.
Indeed, each term of the resulting function would be missing two arguments,
namely $i$ and $j$. Moreover, there will be two terms with the same missing $(i, j)$,
and depending on the order of their removal they will have opposite exponents.
For term, where the greater index was removed first it will be $(-1)^{i+j}$,
in another case it is $(-1)^{i+j-1}$, as when removing the argument with the greater index,
its index is reduced by 1.

There is also the set of cochains, for which $\delta \nu_k = 1$.
Those are called cocycles or closed forms.
It is obvious, that all coboundaries are cocycles.
The structure of action of $\delta$ on $C_k$ spaces is schematically shown on \reff{cohoms}.

It is known, that $k$-cocycle set can be factorized by $k$-coboundary set.
It will be easier to show in the physical implementation, so we skip a proof here.
This Factor space is called $k$-th cohomology of $G \rightarrow F$ mapping
and is denoted as
\begin{equation}
    H^k(G, F) = \frac{k\text{-cocycles}}{k\text{-coboundaries}} =
    {\text{Ker}(\delta_{k+1})}/{\text{Im}(\delta_k)}
\end{equation}

While studying this kind of function it is useful to introduce functions
\begin{equation}
\begin{split}
    \omega_k(g_1, ..., g_n) &= \nu_k(0, g_1, g_1 + g_2, ..., g_1 + g_2 + ... + g_k) \\
    \nu_k(g_0, g_1, ..., g_n) &= \omega_k(-g_0 + g_1, -g_1 + g_2, ..., -g_{k-1} + g_k )
    \label{eq:nu_omega}
\end{split}
\end{equation}
The $\omega_n$ have one less argument and no additional symmetry condition.
One-to-one mapping between $\omega_n$ and
$\nu_n$ (the whole symmetric equivalence class) is then guaranteed.
The downside of this is that the action of the coboundary operator becomes complicated.
With straightforward calculations, one can get
\begin{equation}
\begin{split}
    \delta \omega_k(g_0, g_1, ..., g_k) = \omega_k(g_1, ..., g_k) &\cdot \\ \cdot
    \prod_{i=1}^k \omega_k(g_0, g_1,..., g_{i-2}, &g_{i-1} + g_{i}, g_{i+1}, ..., g_k)^{(-1)^i}
    \cdot \omega_k(g_0, ..., g_{k-1})^{(-1)^{k+1}}
\end{split}
\end{equation}
The precise expressions for smaller values of $k$ are
\begin{equation}
\begin{split}
    \delta \omega_0 (g_0) &= 1 \\
    \delta \omega_1 (g_0, g_1) &= \frac{\omega_1(g_1) \omega_1(g_0)}{\omega_1(g_0 + g_1)} \\
    \delta \omega_2 (g_0, g_1, g_2) &= \frac{\omega_2(g_1, g_2) \omega_2(g_0, g_1 + g_2)}{\omega_2(g_0 + g_1, g_2) \omega_2(g_0, g_1)} \\
    \delta \omega_3 (g_0, g_1, g_2, g_3) &= \frac{\omega_3(g_1, g_2, g_3) \omega_3(g_0, g_1 + g_2, g_3) \omega_3(g_0, g_1, g_2)}{\omega_3(g_0 + g_1, g_2, g_3) \omega_3(g_0, g_1, g_2 + g_3)}
    \label{eq:d_omega}
\end{split}
\end{equation}

\subsection{Unitary transformations}

SPT phases protected by symmetry group $S$ are known
\cite{spt-cohom-3,spt-cohom-4,spt-cohom-5}
to be described by the cohomology groups $H^{d+1}(S, U(1))$,
where $d$ is the system dimension.
Here we will explicitly show how cohomology is involved
in the Hamiltonian construction,
to provide an intuitive understanding odd the situation.  

We will be working on a two-dimensional triangular lattice
(this technique also works for $d$-dimensional lattices,
such as a line in case $d=1$, a tetrahedron lattice for $d=3$, etc.
Some insights about the differences will be provided alongside).
Each node's state is given by an element of additive representation $N$ of group $S$.

Suppose we have our trivial Hamiltonian,
that is just a sum of some commutative elements over the lattice.
This Hamiltonian will be symmetric under any $S_0$,
that is a product of an arbitrary operator $s$ from $S$ over all the lattice.  
Now we need to look for the $S_0$-symmetric local unitary transformations
to construct the whole phase space.
Let's try to write $U$ as a product of $({S} \rightarrow U(1))$ 3-forms 
(or $d+1$-forms for $d$-dimensional case) over triangles as
\begin{equation}
    U = \prod_\Delta U_\Delta^{\epsilon_\Delta} =
    \prod_\Delta \nu_3 (0, n_1^\Delta, n_2^\Delta, n_3^\Delta)^{\epsilon_\Delta}
\end{equation}
where $n_i^\Delta$ are states of particles on the triangle and
$\epsilon_\Delta = \pm1$ indicates the orientation of triangle
based on the direction (clockwise or counterclockwise) of indexing.

\begin{figure}
    \begin{subfigure}[l]{0.4\textwidth}
        \centering
        \includegraphics[width = \textwidth]{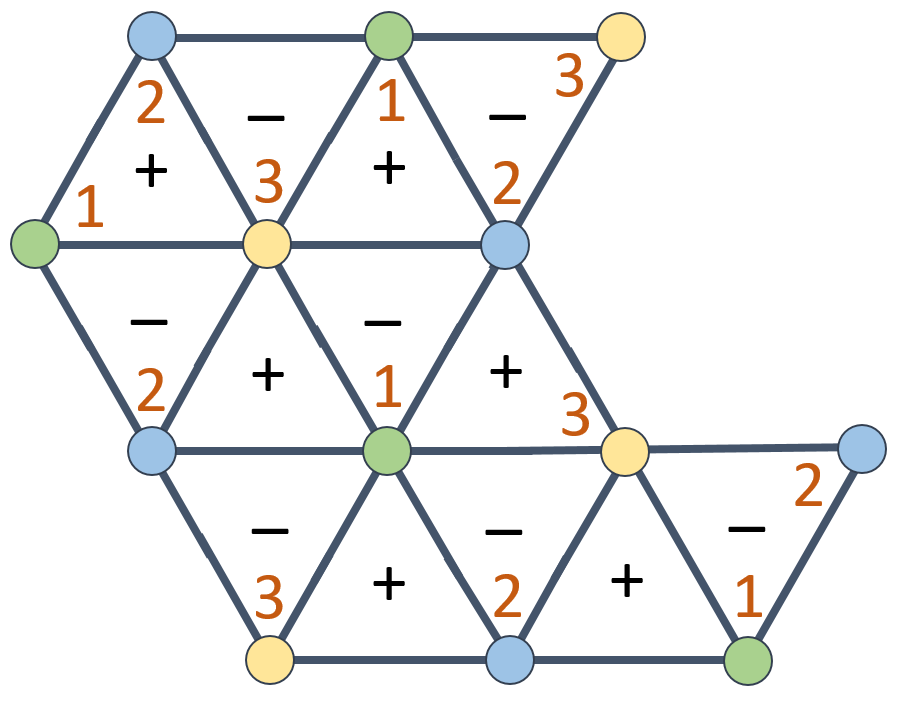}
        \caption{\label{fig:color_index}Color-based indexing.}
    \end{subfigure}
    \hfill
    \begin{subfigure}[l]{0.55\textwidth}
        \centering
        \includegraphics[width = \textwidth]{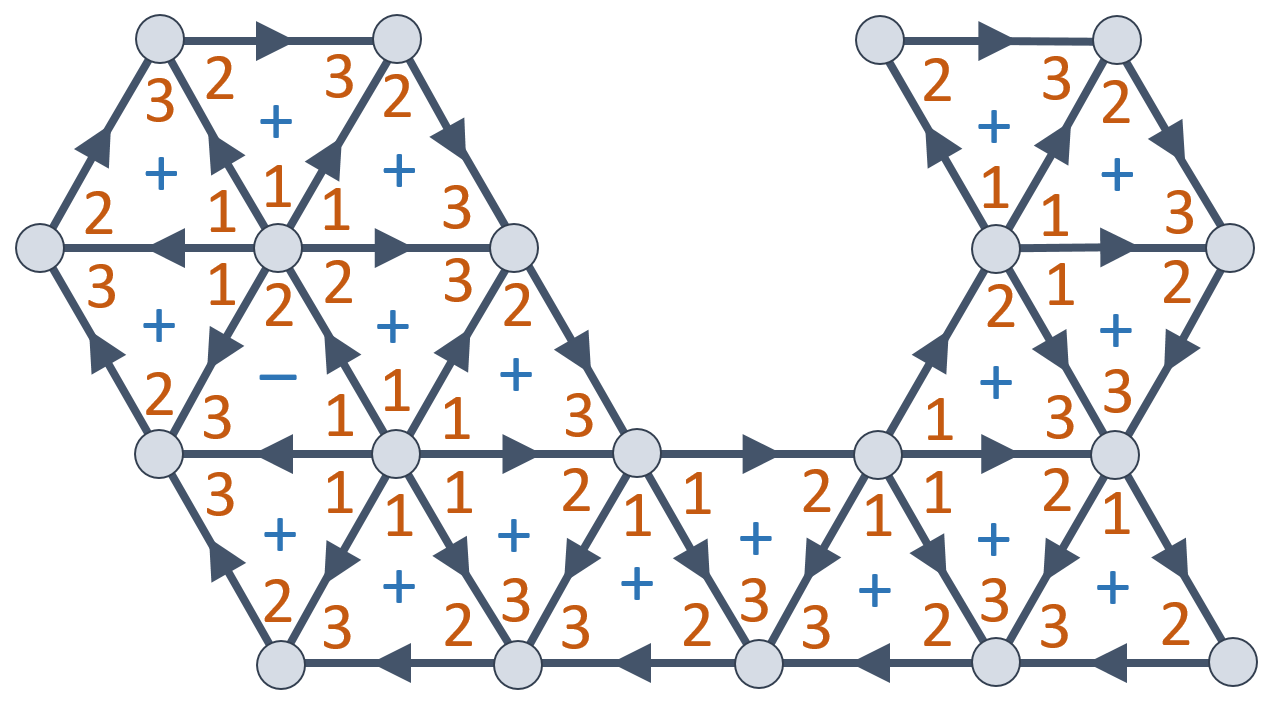}
        \caption{\label{fig:arrow_index}Arrows-based indexing.}
    \end{subfigure}
    \caption{\label{fig:point_index}Indices of points and orientations of triangles
        for different ways of indexing the lattice}
\end{figure}

We will discuss two of the ways of indexing a lattice.
The first one is fairly simple.
The lattice is divided into three larger sub-lattices,
usually denoted with different colors,
and all the nodes of that sub-lattice have the same index,
as shown on \reff{color_index}.

In the other approach, arrows are first drawn on each link of the lattice \reff{arrow_index} \cite{spt-cohom-2, spt-cohom-3}.
The only restriction is that there should be no arrows making a cycle inside one triangle
(or other structure block of the lattice, like tetrahedron in $d=3$, etc.).
Then the indexing is done for each triangle (or other structure block) independently,
in a way that each arrow points from the smaller index to the larger.
This is a one-to-one mapping between arrows and indexing.
Note, that the same node might have different indices in different triangles.

Generally speaking, these transformations are not symmetric,
but the situation changes if we demand for $\nu_3$ to be cocycles.
This condition is written as
\begin{equation}
    \delta\nu_3(n, n', n_1, n_2, n_3) =
    \frac{\nu_3(n', n_1, n_2, n_3) \nu_3(n, n', n_2, n_3) \nu_3(n, n', n_1, n_2)}
    {\nu_3(n, n_1, n_2, n_3) \nu_3(n, n', n_1, n_3)} = 1
\end{equation}
this can be rewritten as
\begin{equation} \label{eq:cycle_nu_symm}
    \frac{\nu_3(n, n_1, n_2, n_3)}{\nu_3(n', n_1, n_2, n_3)} =
    \frac{\nu_3(n, n', n_1, n_2) \nu_3(n, n', n_2, n_3)}{\nu_3(n, n', n_1, n_3)}
\end{equation}
Using this, the symmetry of $\nu$, and denoting $s n s^\dagger = n_s + n$, we will get
\begin{equation} \label{eq:sym_on_U}
    S_0 U_\Delta S_0^\dagger = \nu_3(-n_s, n_1, n_2, n_3) = 
    U_\Delta \frac{\nu_3(0, -n_s, n_1, n_3)}{\nu_3(0, -n_s, n_1, n_2) \nu_3(0, -n_s, n_2, n_3)}
\end{equation}
So $U_\Delta$ is not symmetric under $S$ by itself,
but their product over all the triangles is.
Indeed, in case we do our indexation the color-based way \reff{color_index}
A term $\nu_3(0, -n_s, n_i, n_j)$ will come from two neighboring triangles
that have the $(i, j)$ link in common.
As those triangles always have opposite orientations,
the term appears both in the numerator and the denominator,
and thus vanishes.

The case of arrow-based indexing \reff{arrow_index}
we also consider a link $IJ$ and the two triangles that contain it.
Suppose the arrow points from $I$ to $J$.
The possible index pairs $(i, j)$ of $I$ and $J$ are then
$(1, 2)$, $(2, 3)$ and $(1, 3)$ in both triangles (but not simultaneously).
The triangle left to $I \rightarrow J$ will have
orientations $\inb<->$, $\inb<->$ and $\inb<+>$ correspondingly,
if we denote clockwise by $\inb<+>$.
For the triangle on the right it will be $\inb<+>$, $\inb<+>$, $\inb<->$.
It is worth mentioning once again,
that this doesn't mean always having opposite orientations in neighboring triangles
as the index pair can be different in those triangles.
One may notice, that
for the right triangle, the term $\nu_3(0, -n_s, n_i, n_j)$ will always end up in the numerator,
and similarly, for the left triangle it's always in the denominator.
Thus those terms vanish.

In both cases, we didn't take into account the links that are on the edge of the system
thus don't have their counterparts from a neighboring triangle,
as the symmetry is broken on the edge and this is exactly what we need.
These kinds of transformations are the ones that will be generating our {\Hs}.

Notice, that in the case of color-based indexing,
we are free to modify $\nu_3(0, n_1, n_2, n_3)$
by adding some terms to it that depend only on two of the $n$-s
as they vanish immediately. Let's call those terms $\nu_{3(2)}$,
indicating that formally they depend on three elements, but factually on two.
So we say the set of local symmetric $U$-s is given by
\{3-cocycles\} $\times$ \{non-cocycle $\nu_{3(2)}$-s\}.
Also notice, that in case of the addition of such terms,
the action of symmetry on them will no longer be given by \refe{sym_on_U}
and should be calculated explicitly.

The feature that we get from arrow-based indexing is that
by choosing the arrow configuration
we can make all the links on the edge have the same direction
along the traversal,
make all the edge links to have index pairs $(1, 2)$ or $(2, 3)$
and consequently, all the edge triangles will have the same orientation
as shown in \reff{arrow_index}.
In this case, the source of the local symmetric $U$ set is just \{3-cocycles\}.
In case there are no thin regions
(no edge nodes are connected via a non-edge link)
we can even make all edge links have the same index pair $(1, 2)$ or $(2, 3)$.

\hfill

When we say,
that a Hamiltonian $H_1$ is continuously connected to $H_2$ without breaking symmetry
(let's call these {\Hs} and the corresponding $U$-s equivalent),
this means, that there is a set of symmetric $U^{(\alpha)}, \alpha \in [0, 1]$
continuous on $\alpha$ with $U^{(0)} = 1$ and $U^{(1)} H_1 U^{(1) \dagger} = H_2$.
The $U$-s that we have so far can not be parametrized that way.

The solution is to take $\nu_3$-s to be exact forms.
\begin{equation}
    \nu_3(0, n_1, n_2, n_3) =
    \delta \nu_2(0, n_1, n_2, n_3) =
    \nu_2(n_1, n_2, n_3) \frac{\nu_2(0, n_1, n_3)}
    {\nu_2(0, n_1, n_2) \nu_2(0, n_2, n_3)}
\end{equation}
This has a structure similar to what we have seen in \refe{cycle_nu_symm}.
Using a similar logic one can state,
that only terms $\nu_2(n_1, n_2, n_3)$ will remain after in $U$.
The major advantage over closed forms is that $\nu_2$ has an intrinsic symmetry on $s$,
which means that any $\nu_2(n_1, n_2, n_3)^\alpha$ is also symmetric under $s$.
In their turn, $\nu_2^\alpha$ generate a continuous set of symmetric $U^\alpha$,
where $U^0 = 1$ and $U^1 = U$.

So the set of $U$-s equivalent to $1$ (trivial $U$-s) is given by \{3-coboundaries\}.
In a similar way to what was suggested for just symmetric $U$-s,
in the case of color-based indexing we get an additional
$\times$ \{non-cocycle $\nu_{3(2)}$-s\} here as well.
Any two $U$-s that differ by a trivial $U$ are also obviously equivalent.
So the same-phase space is defined by trivial $U$-s,
and the phases are
\begin{equation}
    \text{\{different phases group\}} =
    \frac{\text{symmetric $U$-s}}{\text{trivial $U$-s}} =
    \frac{\text{3-cocycles}}{\text{3-coboundaries}} =
    H^3(N, U(1))
\end{equation}
In case of $d$ dimensions it would have been $H^{d+1} (N, U(1))$.

Derivation and usage of cohomology groups can be found in
literature \cite{spt-cohom-3, spt-cohom-4, spt-cohom-5, cohoms-1, cohoms-2}

\subsection{Finding a proper $\nu_3$}

A closed but not exact 3-form
will be the source for a Hamiltonian from the non-trivial phase.
To find it, we have to determine a basis for all possible functions.
As it has been mentioned before,
it's the best to look for specific functions in $\omega$-representation
as there are no additional symmetry conditions on them.
Also, it would be helpful if
$\omega$ takes the familiar integers as arguments instead of $S$-elements
(of course, they should be a representation of S).

If group $S$ has $r$ generators $g_{\alpha_1}, g_{\alpha_2}, ..., g_{\alpha_r}$
and $g_{\alpha_i}^{N_i} = 1$, then any element of $S$ can be given as
$s = g_{\alpha_1}^{n_{\alpha_1}} g_{\alpha_2}^{n_{\alpha_2}} ... g_{\alpha_r}^{n_{\alpha_r}}$,
where $n_{\alpha}$-s are some integers.
If we define addition for $n_{\alpha_i}$ to be by $\text{mod } N_i$,
then $\{n_\alpha\}$ (or $\vec{n}$) becomes an additive representation of $S$ in integer numbers.
Keep in mind,
that $\vec{n}_1 + \vec{n}_2$ might be not a component-wise addition
as there can be non-commutative generators in $S$,
which should be taken into account.
For example, if we suppose the permutation group $S_3$
and take $g_{\alpha_1}$ as the generator of rotation
and $g_{\alpha_2}$ as the generator of reflection,
then for the vectors of additive representation, we will have
$(n_1, 0) + (n_2, n') = (n_1 + n_2, n')$
but $(n_1, 1) + (n_2, n') = (n_1 - n_2, 1 + n')$.

The $\omega$ functions basis
(by saying basis we mean that any $\omega$ can be given as a product of these functions)
can be given in a way similar to a simple Taylor series expansion.
The basic monomial functions are
\begin{equation}
    \omega_K^{(c)}(\vec{n}_1, \vec{n}_2, ... , \vec{n}_K) = 
    \exp{\frac{2 \pi i}{N_c} \psi_K^{(c)}(\vec{n}...)}
    \hspace{1cm}
    \psi_K^{(c)}(\vec{n}...) =
    n_{1, \alpha_c}^{b_1} n_{2, \alpha_c}^{b_2} \dots n_{K, \alpha_c}^{b_K}
    \label{eq:basis_omega}
\end{equation}
where $c$ indicates the component of vectors and $b_k$ are some non-negative integers.
As the $n$-s here are integers with no restrictions,
the factor $2 \pi i / N_c$ is there to ensure the equivalence of $n_{k, \alpha_c}$ and $n_{k, \alpha_c} + N_c$
There is only a finite number of these functions,
as for any $N_c$, there is some value of $b \leq N_c$,
starting from which $n^b \equiv n^{b - p_c} (\text{mod } N_c)$, where $p_c > 0$.
It is known as a simple consequence of the Fermat-Euler theorem.
For example $n^4 \equiv n^2 (\text{mod } 4)$,  $n^7 \equiv n (\text{mod } 7)$,
$n^5 \equiv n^3 (\text{mod } 24)$, etc..
The terms of higher order are no new terms.

Then the next step is to calculate the basis for all exact 3-forms.
This is done by simply taking the whole basis of $\omega_2$
and applying $\delta$ as shown in \refe{d_omega} to it.
The produced independent combinations as a whole are our basis.

Afterwards, we need the basis for the closed 3-forms.
Similarly, here $\delta$ should be applied to the whole basis of $\omega_3$
and the combinations that produce $1$ will be the basis.
The part of this basis that is independent of the exact 3-form basis
is the basis for the factor space of 3-cocycles and 3-coboundaries.
Though finding cohomology is a mathematical problem
and we just need one non-exact $\omega_3$
to produce a non-trivial Hamiltonian,
there will be no need to find the whole basis,
but rather a single element in it.

For simplicity of writing, we will be using $\psi$-s instead of $\omega$-s.
Notice that they are defined right up to $\text{mod } N_c$.

\subsection{Construction of a Hamiltonian}

Once we have a non-trivial $\nu_3$ and the corresponding $U$
it needs to be applied to our trivial Hamiltonian $H_0$,
which is composed of local terms $t$ symmetric under $S$.
The initial transformed Hamiltonian would then be written as
\begin{equation}
    H_i = \sum_t U t U^\dagger
\end{equation}
where $t$ runs through all the terms
(there might be single-node terms, two-node terms defined on the links, etc.).

From the previous section, we know
that this transformed Hamiltonian is not symmetric under $S$,
particularly on the edge.
To restore that symmetry we can sum up
all the possible symmetry transformations.
\begin{equation}
    H = \frac{1}{\#S} \sum_{s \in S} \sum_t s U t U^\dagger s^\dagger
    \label{eq:s_sum}
\end{equation}
with $s$ taking all possible values from symmetry group $S$
and $\#S$ being the number of elements in it.
Now this Hamiltonian is explicitly symmetric.
What is left to do is to separate the edge part.

\begin{figure}
    \centering
    \includegraphics[width=250px]{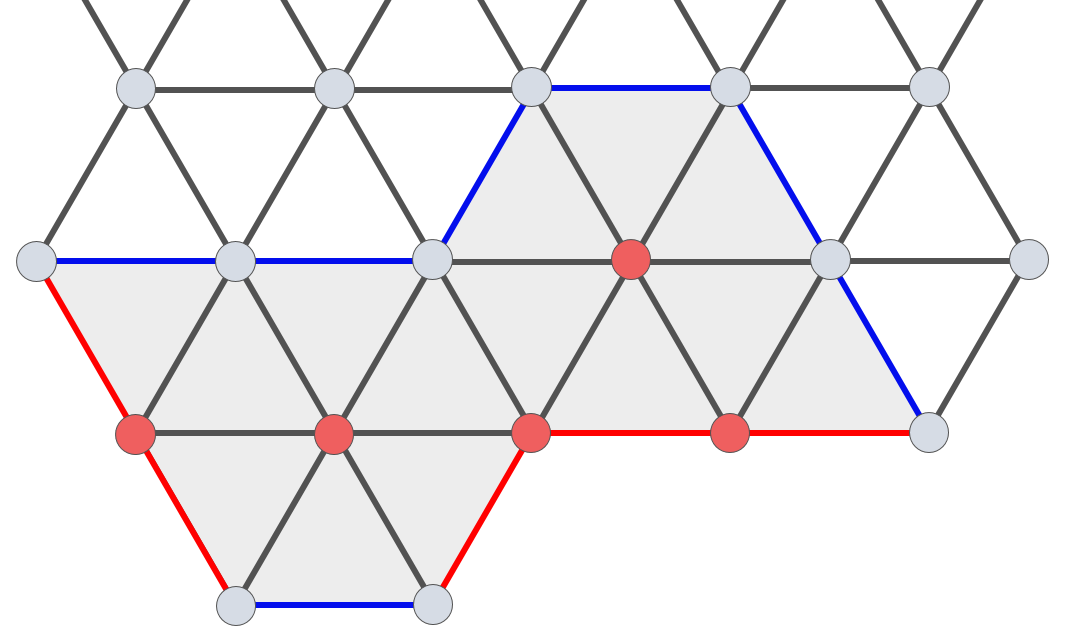}
    \caption{The residual terms generated by action of $s$ on $U t U^\dagger$.
    (In reality one is hardly going to have such a $t$,
    but it's easier to see the principle of $s$'s action here.)
    Nodes in $p_t$ are marked red.
    Shaded triangles are ones with $[U_\Delta, t] \neq 0$ ($U_{p_t}$ triangles).
    The produced residual terms are on the links marked red or blue ($R_{s, t}$ links),
    and residual terms on blue links are commutative with $t$.
    So only the terms on red links remain ($V_{s, t}$ links).}
    \label{fig:residuals}
\end{figure}

As already mentioned, $U$ is not symmetric on the edge of the system,
and produces residual edge terms $\nu_3(0, -n_s, n_i, n_j)^{\pm 1}$
under the action of symmetry $s$,
as shown in \refe{sym_on_U} and the later interpretation.
Here $n_s = s n s^\dagger - n$ and
$n_i$ and $n_j$ are correspond to nodes of links on the edge.

For any term $t$ given on node set $p_t$,
all the $U_\Delta$-s that make up the $U$ will be commutative to it
except the ones that have a common node with $p_t$.
Let's call this "non-commutative with $t$" part of $U$ a $U_{p_t}$.
The residual terms $R_{s, p_t}$
produced of action of $s$ on $U_{p_t}$
will be defined on the links on the border of triangles
that are contained in $U_{p_t}$.
Only part of $R_{s, p_t}$,
that is defined on links that contain a node form $p_t$
(we will call it $V_{s, p_t}$) is not commutative with $t$.
Those will be the system's edge links that contain a node from $p_t$.
All of this is illustrated on \reff{residuals}.
In mathematical formulation
\begin{equation}
    \begin{split}
    s U t U^\dagger s^\dagger =
    s U_{p_t} t U_{p_t}^\dagger s^\dagger =
    U_{p_t} R_{s, p_t} s t s^\dagger R_{s, p_t}^\dagger U_{p_t}^\dagger &= \\
    =U_{p_t} R_{s, p_t} t R_{s, p_t}^\dagger U_{p_t}^\dagger &=
    U_{p_t} V_{s, p_t} t V_{s, p_t}^\dagger U_{p_t}^\dagger =
    V_{s, p_t} \bar{t} V_{s, p_t}^\dagger
    \end{split}
\end{equation}
where $\bar{t} = U t U^\dagger$
Notice that we have done nothing but cancel out terms.
So the $V_{s, p_t}$ terms are exactly the ones shown in \refe{sym_on_U}.
Also as it is seen in \reff{residuals},
$V_{s, p_t}$ will only have terms on links $\inb<{q r}>$
on the edge $\partial$ of the system,
that contains a node from $p_t$.
So the Hamiltonian can be written as
\begin{equation}
    H = \frac{1}{\#S} \sum_t \sum_{s \in S} V_{s, p_t} \bar{t} V_{s, p_t}^\dagger
    \hspace{1cm} V_{s, p_t} = \prod_{
        \substack{\inb<{q r}> \in \partial \\ i_q < i_{r} \\ q \text{ or } r \in p_t}}
    \nu_3(0, -n_s, n_q, n_r) ^ {\epsilon_{q r} (-1)^{i_q + i_r}}
    \label{eq:def_V}
\end{equation}
where $\epsilon_{q r}$ is the orientation of the triangle containing those points,
$i_q$ and $i_r$ are the corresponding indices in that triangle,
and the term $(-1)^{i_q + i_r}$ in the exponent
indicates the initial position (numerator vs denominator) of the $\nu_3$ in \refe{sym_on_U}.

It is obvious, that $V_{s, p_t}$ is $1$ for the $t$ that are fully emerged in the bulk.
\begin{equation}
    H = H_B + H_\partial \hspace{1cm}
    H_B = \sum_{t \in \partial^*} \bar{t} \hspace{1cm}
    H_\partial = \frac{1}{\#S} \sum_{t \in \partial} \sum_{s \in S} V_{s, p_t} \bar{t} V_{s, p_t}^\dagger
\end{equation}
with $\partial^*$ denoting the bulk.
One can notice that each bulk term is commutative with any bulk or edge term,
however, the edge terms are not commutative to each other in general.
So we got a non-trivial edge model that is separated from the bulk.

We might also want to have translational symmetry on the edge,
i.e. ability to write all the components of $V_{s, p_t}$ in \refe{def_V}
without dependence on the point indices.
Let's take a node $p$ and its next node along clockwise traversal $p+1$.
The $q$ in \refe{def_V} is the one with the smaller index of the two.
In the case of arrow-based indexing, there is nothing further to be done
as we will always have $q = p$, $r = p+1$ and
$\epsilon_{q r} (-1)^{i_q + i_r} = -1$.

In case of color-based indexing $\epsilon_{q r}$ is
$1$ if $(i_p, i_{p+1}) \in \{(1, 2), (2, 3), (3, 1)\}$
and is $-1$ otherwise.
$(-1)^{i_q + i_r} = 1$
if $(i_p, i_{p+1}) \in \{(1, 3), (3, 1)\}$
and is $-1$ otherwise.
So overall exponent is $-1$ if $i_p < i_{p+1}$
and is $1$ otherwise (if $i_p > i_{p+1}$).
Our objective here is to make the expression under the product "antisymmetric"
to permutation of the last two arguments as
$f(0, -n_s, n_a, n_b) = f(0, -n_s, n_b, n_a)^{-1}$.
This way the additional condition $i_q < i_r$ under the product can be removed,
because the $-1$ from the exponent and
the $-1$ from anti-symmetry will cancel each other.
Here the freedom to add $\nu_{3(2)}$-s proves helpful.
The addition of a $\nu_{3(2)}$ to the $U_\Delta$
adds a factor $s \nu_{3(2)} s^\dagger / \nu_{3(2)}$ to the right side of \refe{sym_on_U}.
In order not to break the existing symmetry for $i_p < i_{p+1}$ case,
three terms $\nu_{3(2)}^{(1)}(0, n_1, n_2, n_3) = f_2 (n_1, n_2)^{-1}$,
$\nu_{3(2)}^{(2)}(0, n_1, n_2, n_3) = f_2 (n_2, n_3)^{-1}$ and 
$\nu_{3(2)}^{(3)}(0, n_1, n_2, n_3) = f_2 (n_1, n_3)$
should be added at a time.
This way new factors $f_2(n_s + n_i, n_s + n_j) f_2(n_i, n_j)^{-1}$
appear next to corresponding $\nu_3(0, -n_s, n_i, n_j)$.
We can interpret it as modifying $\nu_3$ like
\begin{equation}
    \nu_3(0, -n_s, n_a, n_b) \rightarrow
    \nu_3(0, -n_s, n_a, n_b) \frac{f_2(n_s + n_a, n_s + n_b)}{f_2(n_a, n_b)}
    \label{eq:symm_nu}
\end{equation}
which has a chance to be anti-symmetrized,
as the set of linearly independent $f_2$ elements is larger then
the number of imposed conditions.
Later we will mean the anti-symmetrized version when we refer to $\nu_3$,
which will no longer have an intrinsic symmetry.

Afterwards for both indexing techniques $V_{s, p_t}$ can be written as
\begin{equation}
    V_{s, p_t} = \prod_{\inb<{p, p+1}> \cap p_t }
    \nu_3^{-1}(0, -n_s, n_p, n_{p+1})
    \label{eq:V_sym}
\end{equation}
The product runs through all edge terms that contain any point of $p_t$.

After choosing $\nu_3$,
It is now possible to explicitly write the expression for
an edge Hamiltonian from a non-trivial SPT phase.

\section{SPT models with $Z_2$ paramagnetic Ising model}

Summing up the information in the previous section, the steps for creating the model are:
\begin{enumerate}
    \item Choose a symmetry and the corresponding additive representations
    \item Find a non-exact 3-cocycle (and symmetrize it if needed)
    \item Write the trivial Hamiltonian that has the symmetry (under the lattice product symmetry)
    \item Calculate the transformed edge elements
\end{enumerate}

The simplest application of this algorithm can be seen
on the simplest symmetry group $Z_2$ with only generator $g_{\alpha_1}$, $g_{\alpha_1}^2 = 1$.
We will be using a Hamiltonian that is symmetric under $Z_2$ in a representation
where $g_{\alpha_1}$ is the $\sigma^x$ Pauli matrix.
Then the corresponding additive representation is $n \in \{0, 1\}$, $n = (\sigma^z + 1) / 2$.
Mathematically we know that $H^3(Z_2, U(1)) = Z_2$,
so there should be an SPT phase here.

The only non-exact 3-cocycle is given by $\psi_3(n_1, n_2, n_3) = n_1 n_2 n_3$
which corresponds to $\nu_3(0, n_1, n_2, n_3) = -1^{n_1 n_2 n_3 - n_1 n_3}$.

For the case of color-based indexing, we want it to be antisymmetric
on the permutation of the last two arguments.
It is to be done as mentioned in \refe{symm_nu}.
$f_2$-s that depend on a single argument change nothing in terms of symmetry,
and the only other basic $f_2(n_1, n_2) = -1^{n_1 n_2}$ generates a term
$-1^{n_1 + n_1  n_2 + n_1  n_3}$ next to $\nu_3$.
So adding half (of the exponent) of it will do the trick.
Thereby the expressions for $\nu_3$ for arrow-based and color-based indexings will correspondingly be
\begin{equation}
    \nu_3^{(a)} (0, -n_s, n_a, n_b) = -1^{n_s (n_a - 1) n_b} \hspace{1cm}
    \nu_3^{(c)} (0, -n_s, n_a, n_b) = i^{n_s (2 n_a n_b + n_a - n_b + 1)}
\end{equation}
The $\nu$ is trivial for $n_s = 0$.
If the initial Hamiltonian just consists of elements $\sigma_p^x$
then each one of them on the edge will transform as
\begin{equation}
    \sigma_p^x \rightarrow \frac{1}{2}
    \inb[{\bar{\sigma}_p^x + V_{1, p} \bar{\sigma}_p^x V_{1, p}^\dagger}]
\end{equation}
where $V$ is given by \refe{V_sym}.
We will later drop the factor in front,
as it is some unimportant constant when we consider only the edge.
The bars in the expression explicitly indicate that
the new operators satisfy Pauli matrix algebra just like the old ones,
however, they are given in a different representation.

$V_{1, p}$ will only contain terms of links $\inb<{p-1, p}>$ and $\inb<{p, p+1}>$
and after straightforward calculations, one can get
the final expression of transformed edge $\sigma_p^x$
\begin{equation}
    \sigma_p^x \rightarrow \bar{\sigma}_p^x - \bar{\sigma}_{p-1}^z \bar{\sigma}_p^x \bar{\sigma}_{p+1}^z
    \label{eq:new_sigma}
\end{equation}
where $\bar{\sigma}^z$ is the same as $\sigma^z$.
This is due to $\sigma^z$-s being commutative with $U$
that makes the transformation to barred operators.
Notice that we got the same Hamiltonian for both indexing techniques
despite of having different $\nu_3$-s.

\begin{figure}
    \centering
    \begin{subfigure}[l]{0.5\textwidth}
        \includegraphics[width=\textwidth]{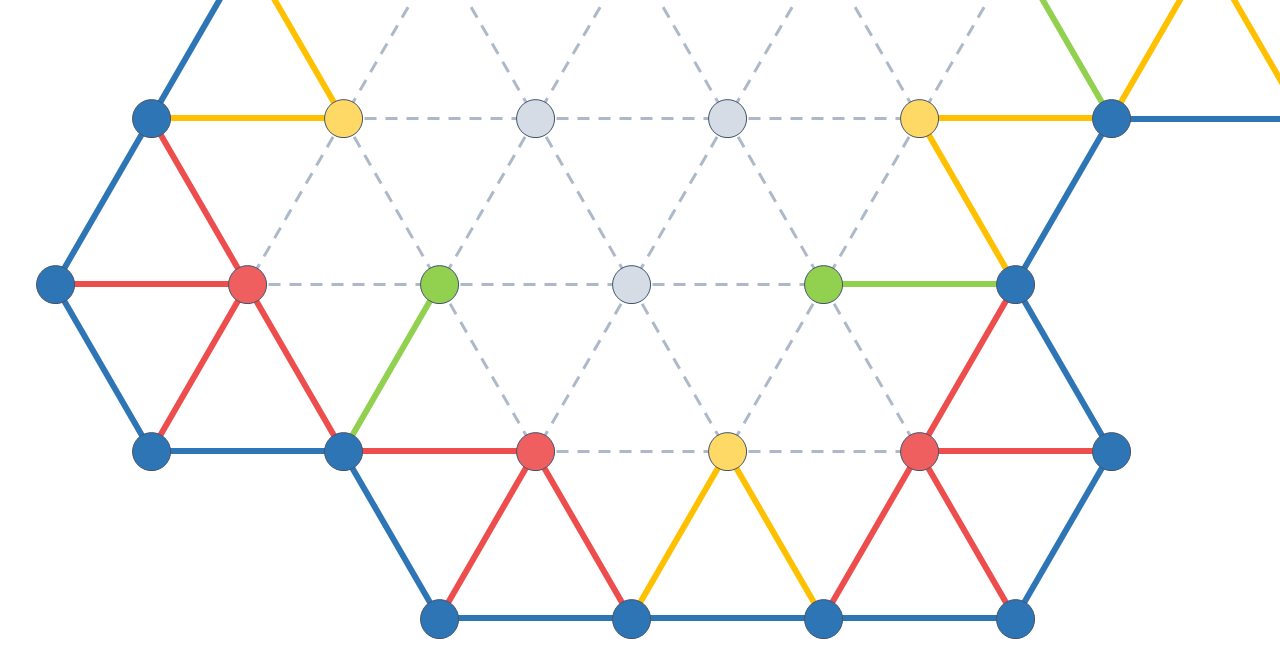}
        \caption{\label{fig:custom_edge_inter}Custom edge.
            The multiple-shared bulk nodes and corresponding links are marked in red
            and the not-shared ones are green.}
    \end{subfigure}
    \begin{subfigure}[l]{0.5\textwidth}
        \includegraphics[width=\textwidth]{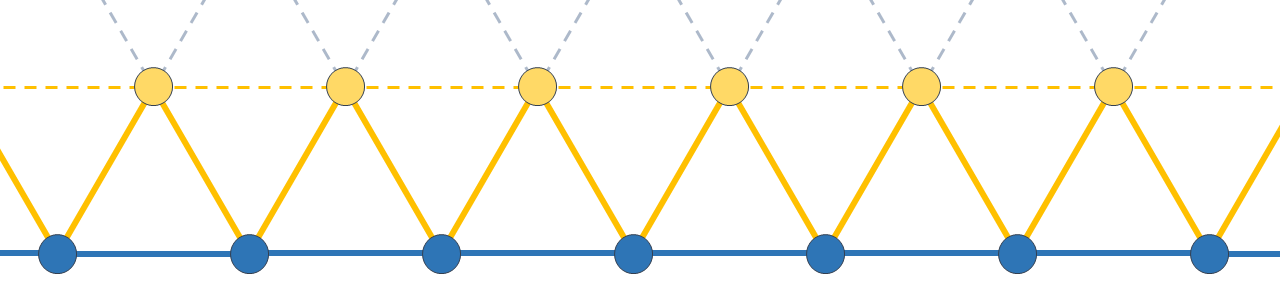}
        \caption{\label{fig:flat_edge_inter}Flat edge.}
    \end{subfigure}
    \caption{\label{fig:edge_inters}The interactions between edge (blue) and edge-neighboring bulk (yellow) nodes
        for different edge configurations.
        Edge-edge and edge-bulk interaction links are colored red and yellow correspondingly.
        The rest is dashed.}
\end{figure}
As it was supposed,
we got ourselves a translation-invariant edge Hamiltonian,
which is written as just a sum of elements in \refe{new_sigma} over the edge.
The same Hamiltonian was derived by Levin and Gu
using a very different approach\cite{levin-gu}
through a rather complicated and arbitrary procedure.

\subsection{The boundary model}

Now that we have the easy concept of SPT model derivation,
we can modify our initial Hamiltonian to a ferromagnetic or antiferromagnetic
$Z_2$ Ising model
\begin{equation}
    H = \sum_\inb<{pq}> \sigma_p^x \sigma_q^x
\end{equation}
where the sum runs through all neighboring nodes $\inb<{pq}>$ of the lattice,
and try to generate a non-trivial SPT phase Hamiltonian for it.
In this case, we will have two kinds of edge terms:
the ones defined on the edge links, and ones with one node on the edge.

For a term defined on edge link $\inb<{p, p+1}>$,
the $V$ will contain three links $\inb<{p-1, p}>$, $\inb<{p, p+1}>$ and $\inb<{p+1, p+2}>$.
It is easy to understand,
that each of the $\sigma^x$-s in the product will transform
the way they did in each term of \refe{new_sigma}.
After a little simplification, we get
\begin{equation}
    \sigma_p^x \sigma_{p+1}^x \rightarrow \bar{\sigma}_p^x \bar{\sigma}_{p+1}^x
    (1 - \bar{\sigma}_{p-1}^z \bar{\sigma}_p^z \bar{\sigma}_{p+1}^z \bar{\sigma}_{p+2}^z)
\end{equation}

The transformation of terms with one edge node $p$ and a bulk node $q$
is even more similar to \refe{new_sigma}, as the bulk term doesn't transform.
\begin{equation}
    \sigma_p^x \sigma_q^x \rightarrow \bar{\sigma}_q^x \bar{\sigma}_p^x (
        1 - \bar{\sigma}_{p-1}^z \bar{\sigma}_{p+1}^z) 
\end{equation}

Thus the edge Hamiltonian can be written as
\begin{equation}
    H_\partial = \sum_{\inb<{p, p+1}> \in \partial} \bar{\sigma}_p^x \bar{\sigma}_{p+1}^x
    (1 - \bar{\sigma}_{p-1}^z \bar{\sigma}_p^z \bar{\sigma}_{p+1}^z \bar{\sigma}_{p+2}^z)
    + \sum_{\substack{\inb<{p, q}>\\ p \in \partial, q \notin \partial}}
    \bar{\sigma}_q^x \bar{\sigma}_p^x
    (1 - \bar{\sigma}_{p-1}^z \bar{\sigma}_{p+1}^z)
    \label{eq:e_ham}
\end{equation}
Here we don't discuss the case of the existence of thin regions in the lattice,
when two edge nodes are connected by a non-edge link.
In that case, there would be other terms as well.

One may notice, that bulk node operators $\bar{\sigma}_q^x$
are commutative with the Hamiltonian and with each other,
so they can be interpreted as an independent classical gauge field on the edge system.

In general, any two neighboring edge nodes will have at least one common bulk neighbor.
A single bulk node can be shared between multiple (more than two) edge nodes
or might be connected to only one edge node.
The situation depends on the curvature of the edge.
This can be seen on \reff{custom_edge_inter}.
In the case of flat edge,
each edge node will have exactly two bulk neighbors
with each of them being shared with either of its edge neighbors
as shown on \reff{flat_edge_inter}.
In this case, the Hamiltonian is translation-invariant.
The flat-edge Hamiltonian can also be written as a sum of triangle terms,
but it will still have the next-to-nearest neighbor interaction.

One can parametrise the Hamiltonian \refe{e_ham} as
\begin{equation}
    H_\partial(\boldsymbol{\lambda}) = \sum_{\inb<{p, p+1}> \in \partial} \bar{\sigma}_p^x \bar{\sigma}_{p+1}^x
    (\lambda_1 - \lambda_2 \prod_{p'=p-1}^{p+2}\bar{\sigma}_{p'}^z)
    + \sum_{\substack{\inb<{p, q}>\\ p \in \partial, q \notin \partial}}
    \bar{\sigma}_q^x \bar{\sigma}_p^x
    (\lambda_3 - \lambda_4\bar{\sigma}_{p-1}^z \bar{\sigma}_{p+1}^z)
\end{equation}
It can be seen, that the $H_\partial = H_\partial(1,1,1,1)$ is self-dual under duality transformation
\begin{equation}
\left.\begin{split}
    \bar{\sigma}_p^x \rightarrow -\bar{\sigma}_{p-1}^z \bar{\sigma}_p^x \bar{\sigma}_{p+1}^z, p \in \partial\\
    \bar{\sigma}_p^z \rightarrow \bar{\sigma}_p^z, p \in \partial\\
    \bar{\sigma}_q^\alpha \rightarrow \bar{\sigma}_q^\alpha, q \notin \partial
\end{split}\right\} \Rightarrow H_\partial(\lambda_1, \lambda_2, \lambda_3, \lambda_4)
\rightarrow H_\partial(\lambda_2, \lambda_1, \lambda_4, \lambda_3)
\end{equation}
It is an indication of gapless spectrum.
The self-duality is a consequence of \refe{s_sum}, and the transformation is given by
the initial symmetry operator $S=\prod_p \sigma_p^x$ (with not barred operators),
or equivalently, by the operator $V_{1, \partial}$.
Another apparent symmetry of the system is $\bar{S}=\prod_{p\in \partial} \bar{\sigma}_p^x$,
which generates a trivial transformation $H_\partial(\boldsymbol{\lambda}) \rightarrow H_\partial(\boldsymbol{\lambda})$.

The third global $Z_2$ symmetry of the model is specific to this interacting case
and doesn't appear in the paramagnetic case. It is given by $Z=\prod_p \sigma_p^z$.
Note that $[V_{1,\partial}, \bar{S}]=[V_{1,\partial}, Z]=0$ and
$[Z, \bar{S}]_\pm = 0$ where the "$\pm$" in the index specifies
an anti-commutator or a commutator depending on the parity of the overall node number.
This brings the whole symmetry group to $Z_2 \ltimes (Z_2 \times Z_2)$.

The system also has a hidden symmetry of the "domain wall" number as in \cite{levin-gu}, given by
\begin{equation}
    N=\sum_{p\in\partial} w_{p, p+1} \quad , \quad
    w_{p, p+1} = \frac{1-\bar{\sigma}_p^z \bar{\sigma}_{p+1}^z}{2}
\end{equation}
where $w_{p, p+1}$ is $1$ when $p$ and $p+1$ belong to different domains of
$\inb|+>$ and $\inb|->$ states (eigenstates of $\bar{\sigma}^z$), and is $0$ otherwise.
The relation $[H_\partial, N]=0$ can be checked directly, however, it is easier to see,
that different terms of the Hamiltonian acting on eigenstates of $\bar{\sigma}^z$
allow only local transitions of types
\begin{equation}
\begin{split}
    \inb|{\cdots, \pm, +, -, \pm, \cdots}> &\rightleftarrows \inb|{\cdots, \pm, -, +, \pm, \cdots}>\\
    \inb|{\cdots, \pm, +, +, \mp, \cdots}> &\rightleftarrows \inb|{\cdots, \pm, -, -, \mp, \cdots}>\\
    \inb|{\cdots, \pm, +, \mp, \cdots}> &\rightleftarrows \inb|{\cdots, \pm, -, \mp, \cdots}>\\
\end{split}
\end{equation}
which conserve $N$ locally.

This symmetry can be also presented as a discrete continuity equation
of some two-component current $j^\mu = (w, m)$ as
\begin{equation}
    \partial_\mu j^\mu = \partial_t w - \nabla_p m = 0 \equiv
    i [H, w_{p, p+1}] - (m_{p+1} - m_p) = 0
\end{equation}
A straightforward calculation leads to the expression
\begin{equation}
    m_p = \sigma_p^x (\sigma_{p+1}^y \sigma_{p+2}^z - \sigma_{p-1}^y \sigma_{p-2}^z)
    + \sigma_p^y (\sigma_{p-1}^x \sigma_{p+1}^z - \sigma_{p-1}^z \sigma_{p+1}^y)
    + A_p \sigma_p^y (\sigma_{p+1}^z - \sigma_{p-1}^z)
\end{equation}
for $m_p$, where $A_p$ stand for
the sum of all bulk $\bar{\sigma}_q^x$-s adjacent to $p$.
This expression is useful for a consequent numerical search
of possible holomorphic currents and Kac-Moody algebra in the thermodynamic limit,
which provides information about the conformal field theory describing the boundary \cite{our-1,our-2}.

The Hamiltonian can be brought to a simpler form by introducing
a set of non-local operators
\begin{equation}
    \tau_p^x = \prod_{p'\leq p} \bar{\sigma}_p^x \quad,\quad
    \tau_p^z = \bar{\sigma}_p^z \bar{\sigma}_{p+1}^z
\end{equation}
which also satisfies Pauli matrix algebra. The Hamiltonian takes the form
\begin{equation}
    H_\partial = \sum_{p\in\partial}
    (\tau_p^x \tau_{p+2}^x + \tau_p^y \tau_{p+2}^y) +
    \sum_{p\in\partial} A_p (\tau_p^x \tau_{p+1}^x + \tau_p^y \tau_{p+1}^y)
\end{equation}
The issue of vague definitions of the transformations
due to periodic boundary conditions can be addressed via the introduction
of an additional gauge field, which results in nothing but a need to consider
both periodic and anti-periodic boundary conditions in the emerged model \cite{levin-gu}.

Through further fermionization by a Jordan-Wigner transformation given by
$c_p^\pm = \tau_p^\pm \prod_{p'<p} \tau_{p'}^z$, where
$\tau_p^\pm = \tau_p^x \pm i \tau_p^y$ brings the Hamiltonian to form
\begin{equation}
    H_\partial / 2 = \sum_{p\in\partial} (-1)^{c^+_p c_p} (c^+_{p-1} c_{p+1} + \text{h.c.})
    + \sum_{p\in\partial} A_p (c^+_{p} c_{p+1} + \text{h.c.})
\end{equation}
where the relation $\tau_p^z = 2 c^+_p c_p - 1 = (-1)^{c_p c^+_p}$ is used.
In these notations, the domain-wall symmetry obtains a simple form
$N=L-\sum_p c^+_p c_p$, where $L$ is the size of the system.

\subsection{The 't Hooft anomaly}

SPT phases are known \cite{spt-cohom-4, thooft-2} to be characterized
by the anomalous symmetries on the boundary.
Particularly during gauging, the symmetry group is preserved 
only in its projective representation ('t Hooft anomaly).
This phenomenon can be directly observed
once the symmetries are considered on a finite chain \cite{our-1, our-2}.
Here additional phase factors appear in group product relations,
emphasizing the emergence of a projective representation.

Our symmetry group of interest $Z_2 \ltimes (Z_2 \times Z_2)$ consists of
generators $Z, V_{1,\partial}$ and $\bar{S}$.
We introduce reduced operators on a semi-infinite chain
\begin{equation}
    Z_r = \prod_{p=0}^{\infty} \bar{\sigma}_p^z \quad,\quad
    V_{1,r} = \prod_{p=0}^{\infty} i^{2 n_p n_{p+1} + n_p - n_{p+1} + 1}
    \quad,\quad \bar{S}_r = \prod_{p=0}^{\infty} \bar{\sigma}_p^x
\end{equation}
The anomaly then can be seen as a broken associativity condition for the reduced group.
Using notation $G(\boldsymbol{g}) = Z_r^{g_1} V_{1,r}^{g_2} \bar{S}_r^{g_3}$
where $g_i \in \{0, 1\}$, one can check that
\begin{equation}
    G(\boldsymbol{g}) \inb({G(\boldsymbol{h}) G(\boldsymbol{k})})
    = \omega(\boldsymbol{g}, \boldsymbol{h}, \boldsymbol{k})
    \inb({G(\boldsymbol{g}) G(\boldsymbol{h})}) G(\boldsymbol{k})
\end{equation}
with a nontrivial phase factor
$\omega(\boldsymbol{g}, \boldsymbol{h}, \boldsymbol{k})=(-1)^{g_3 h_3 k_2}$.
As no "$1$" indices appear,
the $Z_r$ generator doesn't contribute to the anomaly.

\section{Conclusion}
We generalized the previously known ideas \cite{levin-gu,our-1,our-2}
and presented a technique for creating SPT phase models
as extensions on any two-dimensional systems
with arbitrary non-trivial symmetry $S$.
The technique relies on the explicit application of
specially defined symmetry-restored quasi-unitary transformations
on the initial Hamiltonian.
The transformations are based on $H^3(S, U(1))$,
and define the phase space of $S$-symmetric SPT models,
which is also given by $H^3(S, U(1))$\cite{spt-cohom-3}.
The technique guarantees the creation of translation-invariant edge models
once applied on non-interacting systems.

A $Z_2$ SPT model is then constructed
based on the $Z_2$ quantum Ising model in its ferromagnetic phase
by applying this technique with $H^3(Z_2, U(1)) = Z_2$.
A translation-invariant edge model is obtained,
regardless of the base model being an interacting one.
The edge model is self-dual, which indicates a gapless spectrum.
It is also coupled to a classical gauge field.
Other symmetry properties and their anomalies are also studied.
Further research can be done to investigate the physical properties
of the resulting model such as edge conductivity,
finite-size scaling behavior and central charge.

\subsection*{Acknowledgements}

I am grateful to V. Iugov, A. Sedrakyan, T. Hakobyan and T. Sedrakyan
for the valuable discussions.
The research was supported by 
Armenian SCS grants Nos. 20TTAT-QTa009 and 21AG-1C024.

\bibliography{refs.bib}

\end{document}